\documentclass[aps,prd,twocolumn,showpacs,amsmath,amssymb]{revtex4-1}
\usepackage{amsmath} \usepackage{graphicx} \usepackage{subfigure}
\usepackage{epstopdf} \usepackage{color} \usepackage{multirow}
\usepackage{setspace} \usepackage{overpic} \usepackage{amssymb}
\usepackage{ulem}
\usepackage[bookmarksnumbered, pdfstartview=FitH,colorlinks,urlcolor=blue, citecolor=blue,linkcolor=blue] {hyperref}
\usepackage{lineno}
\usepackage{bm}
\usepackage{rotating}
\usepackage{xcolor}
\usepackage{makecell}
\usepackage{mathtext}
\usepackage{mathrsfs}
\usepackage[utf8]{inputenc}

\let\oldequation\equation
\let\oldendequation\endequation
\renewenvironment{equation}
 {\linenomathNonumbers\oldequation}
 {\oldendequation\endlinenomath}

\begin{document}
\definecolor{boslv}{rgb}{0.0, 0.65, 0.58}
\definecolor{Munsell}{HTML}{00A877}
\newcommand{\psip}{\psi^{'}}
\newcommand{\psipp}{\psi(3686)}

\newcommand{\Br}{\mathcal{B}}
\newcommand{\too}{\rightarrow}
\newcommand{\del}{\color{red}\sout}
\newcommand{\new}{\color{blue}\uwave}

\newcommand{\afs}{\alpha_s}
\newcommand{\bgp}{\beta\gamma}
\newcommand{\eff}{\varepsilon}
\newcommand{\sintht}{\sin{\theta}}
\newcommand{\costht}{\cos{\theta}}
\newcommand{\dedx}{dE/dx}

\newcommand{\probfc}{Prob_{\chi^2}}
\newcommand{\probpi}{Prob_{\pi}}
\newcommand{\probka}{Prob_{K}}
\newcommand{\probpr}{Prob_{p}}
\newcommand{\proball}{Prob_{all}}

\newcommand{\chicJ}{\chi_{cJ}}
\newcommand{\gchicJ}{\gamma\chi_{cJ}}
\newcommand{\gchica}{\gamma\chi_{c0}}
\newcommand{\gchicb}{\gamma\chi_{c1}}
\newcommand{\gchicc}{\gamma\chi_{c2}}
\newcommand{\hc}{h_c(^1p_1)}
\newcommand{\qqb}{q\bar{q}}
\newcommand{\uub}{u\bar{u}}
\newcommand{\ddb}{d\bar{d}}
\newcommand{\SSB}{\Sigma^+\bar{\Sigma}^-}
\newcommand{\LLB}{\Lambda\bar{\Lambda}}
\newcommand{\ccb}{c\bar{c}}

\newcommand{\psipto}{\psi^{\prime}\rightarrow \pi^+\pi^- J/\psi}
\newcommand{\ptomm}{J/\psi\rightarrow \mu^+\mu^-}
\newcommand{\ppp}{\pi^+\pi^- \pi^0}
\newcommand{\pip}{\pi^+}
\newcommand{\pim}{\pi^-}
\newcommand{\kap}{K^+}
\newcommand{\kam}{K^-}
\newcommand{\ks}{K^0_s}
\newcommand{\pbar}{\bar{p}}
\newcommand{\jp}{J/\psi\rightarrow \gamma\pi^0}
\newcommand{\je}{J/\psi\rightarrow \gamma\eta}
\newcommand{\jep}{J/\psi\rightarrow \gamma\eta^{\prime}}

\newcommand{\LL}{\ell^+\ell^-}
\newcommand{\EE}{e^+e^-}
\newcommand{\MM}{\mu^+\mu^-}
\newcommand{\GG}{\gamma\gamma}
\newcommand{\TT}{\tau^+\tau^-}
\newcommand{\pp}{\pi^+\pi^-}
\newcommand{\kk}{K^+K^-}
\newcommand{\ppb}{p\bar{p}}
\newcommand{\gpp}{\gamma \pi^+\pi^-}
\newcommand{\gkk}{\gamma K^+K^-}
\newcommand{\gppb}{\gamma p\bar{p}}
\newcommand{\ggee}{\gamma\gamma e^+e^-}
\newcommand{\gguu}{\gamma\gamma\mu^+\mu^-}
\newcommand{\ggll}{\gamma\gamma l^+l^-}
\newcommand{\ppee}{\pi^+\pi^- e^+e^-}
\newcommand{\ppuu}{\pi^+\pi^-\mu^+\mu^-}
\newcommand{\etap}{\eta^{\prime}}
\newcommand{\gpi}{\gamma\pi^0}
\newcommand{\geta}{\gamma\eta}
\newcommand{\getap}{\gamma\etap}
\newcommand{\pppp}{\pi^+\pi^-\pi^+\pi^-}
\newcommand{\ppkk}{\pi^+\pi^-K^+K^-}
\newcommand{\pppr}{\pi^+\pi^-p\bar{p}}
\newcommand{\kkkk}{K^+K^-K^+K^-}
\newcommand{\kskp}{K^0_s K^+ \pi^- + c.c.}
\newcommand{\ppkp}{\pi^+\pi^-K^+ \pi^- + c.c.}
\newcommand{\ksks}{K^0_s K^0_s}
\newcommand{\dphi}{\phi\phi}
\newcommand{\phikk}{\phi K^+K^-}
\newcommand{\ppeta}{\pi^+\pi^-\eta}
\newcommand{\gpppp}{\gamma \pi^+\pi^-\pi^+\pi^-}
\newcommand{\gppkk}{\gamma \pi^+\pi^-K^+K^-}
\newcommand{\gpppr}{\gamma \pi^+\pi^-p\bar{p}}
\newcommand{\gkkkk}{\gamma K^+K^-K^+K^-}
\newcommand{\gkskp}{\gamma K^0_s K^+ \pi^- + c.c.}
\newcommand{\gppkp}{\gamma \pi^+\pi^-K^+ \pi^- + c.c.}
\newcommand{\gksks}{\gamma K^0_s K^0_s}
\newcommand{\gphiphi}{\gamma \phi\phi}

\newcommand{\tpp}{3(\pi^+\pi^-)}
\newcommand{\tppkk}{2(\pi^+\pi^-)(K^+K^-)}
\newcommand{\pptkk}{(\pi^+\pi^-)2(K^+K^-)}
\newcommand{\tkk}{3(K^+K^-)}
\newcommand{\gtpp}{\gamma 3(\pi^+\pi^-)}
\newcommand{\gtppkk}{\gamma 2(\pi^+\pi^-)(K^+K^-)}
\newcommand{\gpptkk}{\gamma (\pi^+\pi^-)2(K^+K^-)}
\newcommand{\gtkk}{\gamma 3(K^+K^-)}

\newcommand{\psp}{\psi(3686)}
\newcommand{\jpsi}{J/\psi}
\newcommand{\ar}{\rightarrow}
\newcommand{\lra}{\longrightarrow}
\newcommand{\jpsito}{J/\psi \rightarrow }
\newcommand{\ptoppjp}{J/\psi \rightarrow\pi^+\pi^- J/\psi}
\newcommand{\pspto}{\psi^\prime \rightarrow }
\newcommand{\ptop}{\psi'\rightarrow\pi^0 J/\psi}
\newcommand{\ptoeta}{\psi'\rightarrow\eta J/\psi}
\newcommand{\ecto}{\eta_c \rightarrow }
\newcommand{\ecpto}{\eta_c^\prime \rightarrow }
\newcommand{\xto}{X(3594) \rightarrow }
\newcommand{\chicJto}{\chi_{cJ} \rightarrow }
\newcommand{\chiczto}{\chi_{c0} \rightarrow }
\newcommand{\chicoto}{\chi_{c1} \rightarrow }
\newcommand{\chictto}{\chi_{c2} \rightarrow }
\newcommand{\pspp}{\psi^{\prime\prime}}
\newcommand{\ptochic}{\psi(2S)\ar \gamma\chi_{c1,2}}
\newcommand{\ppjpsi}{\pi^0\pi^0 J/\psi}
\newcommand{\utoeta}{\Upsilon^{\prime}\ar\eta\Upsilon}
\newcommand{\ww}{\omega\omega}
\newcommand{\wf}{\omega\phi}
\newcommand{\ff}{\phi\phi}
\newcommand{\npsp}{N_{\psp}}
\newcommand{\llb}{\Lambda\bar{\Lambda}}
\newcommand{\llbpi}{\llb\pi^0}
\newcommand{\llbeta}{\llb\eta}
\newcommand{\ppi}{p\pi^-}
\newcommand{\pbpi}{\bar{p}\pi^+}
\newcommand{\lamb}{\bar{\Lambda}}
\def\ctup#1{$^{\cite{#1}}$}
\newcommand{\bfg}{\begin{figure}}
\newcommand{\efg}{\end{figure}}
\newcommand{\bitm}{\begin{itemize}}
\newcommand{\eitm}{\end{itemize}}
\newcommand{\bnum}{\begin{enumerate}}
\newcommand{\enum}{\end{enumerate}}
\newcommand{\btbl}{\begin{table}}
\newcommand{\etbl}{\end{table}}
\newcommand{\btbu}{\begin{tabular}}
\newcommand{\etbu}{\end{tabular}}
\newcommand{\bcl}{\begin{center}}
\newcommand{\ecl}{\end{center}}
\newcommand{\bbt}{\bibitem}
\newcommand{\beq}{\begin{equation}}
\newcommand{\eeq}{\end{equation}}
\newcommand{\beqr}{\begin{eqnarray}}
\newcommand{\eeqr}{\end{eqnarray}}
\newcommand{\red}{\color{red}}
\newcommand{\blue}{\color{blue}}
\newcommand{\yellow}{\color{yellow}}
\newcommand{\green}{\color{green}}
\newcommand{\purple}{\color{purple}}
\newcommand{\brown}{\color{brown}}
\newcommand{\black}{\color{black}}

\title{\boldmath Study of the decays $\chi_{cJ} \to\Lambda \bar{\Lambda} \phi$ }

\author{
\begin{small}
\begin{center}
M.~Ablikim$^{1}$, M.~N.~Achasov$^{4,c}$, P.~Adlarson$^{75}$, O.~Afedulidis$^{3}$, X.~C.~Ai$^{80}$, R.~Aliberti$^{35}$, A.~Amoroso$^{74A,74C}$, Q.~An$^{71,58,a}$, Y.~Bai$^{57}$, O.~Bakina$^{36}$
% ...
\\
\vspace{0.2cm}
(BESIII Collaboration)\\
\vspace{0.2cm} {\it
$^{1}$ Institute of High Energy Physics, Beijing 100049, People's Republic of China\\
% ...
$^{3}$ Bochum  Ruhr-University, D-44780 Bochum, Germany\\
$^{4}$ Budker Institute of Nuclear Physics SB RAS (BINP), Novosibirsk 630090, Russia\\
% ...
$^{35}$ Johannes Gutenberg University of Mainz, Johann-Joachim-Becher-Weg 45, D-55099 Mainz, Germany\\
$^{36}$ Joint Institute for Nuclear Research, 141980 Dubna, Moscow region, Russia\\
% ...
$^{57}$ Southeast University, Nanjing 211100, People's Republic of China\\
$^{58}$ State Key Laboratory of Particle Detection and Electronics, Beijing 100049, Hefei 230026, People's Republic of China\\
% ...
$^{71}$ University of Science and Technology of China, Hefei 230026, People's Republic of China\\
% ...
$^{74}$ University of Turin and INFN, (A)University of Turin, I-10125, Turin, Italy; (B)University of Eastern Piedmont, I-15121, Alessandria, Italy; (C)INFN, I-10125, Turin, Italy\\
$^{75}$ Uppsala University, Box 516, SE-75120 Uppsala, Sweden\\
% ...
$^{80}$ Zhengzhou University, Zhengzhou 450001, People's Republic of China\\
\vspace{0.2cm}
$^{a}$ Deceased\\
% ...
$^{c}$ Also at the Novosibirsk State University, Novosibirsk, 630090, Russia\\
% ...
}\end{center}
\vspace{0.4cm}
\end{small}
}

\date{\today}


\begin{abstract}
   Based on $(2712.4 \pm 14.3) \times 10^{6}$ $ e^{+}e^{-}\to\psi(3686)$ events collected with the BESIII detector operating at the BEPCII collider, we report the first evidence of $\chi_{c0}\to \Lambda\bar \Lambda \phi$ decays and the first  observation of $\chi_{c1,2}\to \Lambda\bar \Lambda \phi$ decays, with significances of $4.5\sigma$, $11.3\sigma$ and $13.0\sigma$, respectively. The decay branching fractions of $\chi_{c0,1,2}\to \Lambda\bar \Lambda\phi$ are measured to be $( 2.99\pm1.24\pm0.19) \times 10^{-5}$, $(6.01\pm0.90\pm0.40 )\times 10^{-5}$, and $(7.13\pm0.81\pm0.36) \times 10^{-5}$, where the first uncertainties are statistical and the second systematic. No obvious enhancement near the $\Lambda \bar{\Lambda}$ production threshold or excited $\Lambda$ state is found in the $\Lambda \phi$ (or $\bar{\Lambda}\phi$) system.
\end{abstract}


\maketitle
\section{Introduction}
In the quark model, the $\chi_{cJ}(J=0,1,2)$ mesons are identified as $^3P_J$ charmonium states.  Due to the conservation of parity, it was long considered impossible for $e^+e^-$ annihilation to directly produce them. As a result, the decays of $\chi_{cJ}$ have not been studied as extensively as the vector charmonium states $J/\psi$ and $\psi(3686)$ in both experiment and theory.
However, $\chi_{cJ}$ mesons can be produced via radiative decays of the $\psi(3686)$ with a sizable branching fraction (BF) of about 9\%~\cite{pdg2022} for each $\chi_{cJ}$ state, thereby offering an ideal environment to investigate their properties and decays.

Studies of the processes involving the $B\bar{B}V$ final states, where $B$ and $V$ denote baryons and vector mesons, respectively, are valuable in the search for possible $B\bar B$ threshold enhancements and excited baryon states decaying into $BV$.  A threshold enhancement in the  $p\bar p$ system
was first observed in the $J/\psi\to \gamma p\bar p$ decay at BESII~\cite{BES:2003aic} and was later confirmed by BESIII with  improved precision~\cite{BESIII:2011aa}.  Subsequently, an enhancement around the $\Lambda\bar{\Lambda}$ production threshold was observed in various processes, such as $e^+e^-\to \phi\Lambda\bar{\Lambda}$~\cite{LLbarphi},
which disfavored an interpretation of the enhancement as originating from the $\eta(2225)\to\Lambda\bar{\Lambda}$ decay~\cite{LLbar_theory1}. In addition, the excited state $\Lambda(1670)$ was observed in the near-threshold reaction $K^-p\to
\Lambda\eta$~\cite{Leta} and in the $\Lambda \eta$
mass spectra in the charmonium decay of $\psi(3686)\to
\Lambda\bar \Lambda \eta$~\cite{abc}. However, experimental results
on the $\Lambda\bar{\Lambda}$ production threshold enhancement and on
excited $\Lambda$ states decaying into $\Lambda \phi$ are still
limited. Comprehensive investigations of the $B\bar{B}V$ system in a wide range of charmonium decays are desirable. So far, only a few
studies of $P$ wave charmonium decays, $\chi_{cJ}\to B\bar{B}V$, have been
performed~\cite{pdg2022,three_body_A,three_body_B,three_body_C}. Search for similar decay modes may provide an opportunity
to further investigate the potential enhancement at the $\Lambda\bar \Lambda$
production threshold and the nature of excited $\Lambda$ states.

In this paper, by analyzing  $(2712.4\pm14.3)\times10^{6}$ $\psi(3686)$  events~\cite{psip_num_2021} collected with the BESIII detector, we present the first experimental studies of $\chi_{cJ}\to\Lambda\bar \Lambda\phi$ decays.
\section{BESIII DETECTOR AND MONTE CARLO SIMULATION}
\label{sec:BES}

  The BESIII detector~\cite{Ablikim:2009aa} records $e^+ e^-$ collisions provided by the BEPCII storage ring~\cite{CXYu_bes3}. The cylindrical core of the BESIII detector covers 93\% of the full solid angle and consists of a helium-based multilayer drift chamber (MDC), a plastic scintillator time-of-flight system (TOF), and a CsI(Tl) electromagnetic calorimeter (EMC), which are all enclosed in a superconducting solenoidal magnet providing a 1.0~T magnetic field. The magnet is supported
by an octagonal flux-return yoke with modules of resistive
plate muon counters (MUC) interleaved with steel. The charged-particle momentum resolution at 1~GeV/$c$ is 0.5\%, and the  d$E/$d$x$ resolution is 6\% for the electrons from Bhabha scattering at 1~GeV. The EMC measures photon energy with a resolution of 2.5\% (5\%) at 1~GeV in the barrel (end-cap) region. The time resolution of the TOF barrel part is 68~ps, while that of the end-cap part is 110~ps. The end cap TOF system was upgraded in 2015 using multigap resistive plate chamber technology, providing a time resolution of 60 ps, which  benefits 86\% of the data used in this analysis~\cite{tof_a,tof_b,tof_c}.

Monte Carlo (MC) simulated data samples are produced with a {\sc geant4}-based~\cite{geant4} software package, which includes the geometric description~\cite{detvis} of the BESIII detector and the detector response. They are used to optimize the event selection criteria, estimate the signal efficiency and the level of background. The simulation models the beam-energy spread and initial-state radiation in the $e^+e^-$ annihilation using the generator {\sc kkmc}~\cite{kkmc_a,kkmc_b}. The inclusive MC sample includes the production of the $\psi(3686)$ resonance, the initial-state radiation production of the $J/\psi$ meson, and the continuum processes incorporated in {\sc kkmc}. Particle decays are generated by {\sc evtgen}~\cite{evtgen_a,evtgen_b} for the known decay modes with BFs taken from the Particle Data Group  (PDG)~\cite{pdg2022} and {\sc lundcharm}~\cite{lundcharm_a,lundcharm_b} for the unknown ones.  Radiation from charged final-state particles is included using the {\sc photos} package~\cite{photos}.

To determine the detection efficiency for each signal process, signal MC samples are generated with a modified data-driven generator BODY3~\cite{evtgen_a,evtgen_b}, which simulates contributions from different intermediate states in data for a given three-body final state, as discussed in Sec.~\ref{Sec:BR_determined}.

\section{EVENT SELECTION}
\label{sec:selection}

The $\Lambda$/$\bar \Lambda$ and $\phi$ particles are reconstructed via their decays $\Lambda/\bar{\Lambda}\to p\pi^-/\bar{p}\pi^+$ and $\phi\to K^+K^-$. Charged tracks detected in the MDC are required to be within a polar angle  $\theta$ such that $|\rm{cos}\theta|<0.93$, where $\theta$ is defined with respect to the symmetry axis of the MDC. The $\Lambda$ and $\bar{\Lambda}$ candidates are reconstructed by combining pairs of oppositely charged tracks with pion and proton mass hypotheses satisfying a secondary vertex constraint~\cite{vertex_second}. Events with at least one $p\pi^{-}(\Lambda)$ and one $\bar{p}\pi^{+}(\bar{\Lambda})$ candidate are selected.  In the case of multiple $\LLB$ pair candidates, the one with the minimum value of $\sqrt{(M_{p\pi^{-}}-m_{\Lambda})^{2}+(M_{\bar{p}\pi^{+}}-m_{\bar{\Lambda}})^{2} }$ is chosen, where $M_{p\pi^-}$ ($M_{\bar{p} \pi^+}$) is the invariant mass of the $p\pi^-$ ($\bar{p}\pi^+$) system, and $m_\Lambda$ ($m_{\bar{\Lambda}}$) is the know mass of $\Lambda$($\bar{\Lambda}$)~\cite{pdg2022}. Considering that $\Lambda(\bar{\Lambda})$ has a relatively long lifetime, we require the decay length of $\Lambda(\bar{\Lambda})$ to be greater than zero. The charged tracks other than those originating from the $\LLB$ pair are taken as originating from the $\phi$ decay.
For these tracks, the distance of closest approach to the $e^{+}e^{-}$ interaction point must be less than 10~cm in the beam axis and less than 1~cm in the plane perpendicular to this axis. The measurements of the flight time in the TOF and ${\rm d}E/{\rm d}x$ in the MDC for each charged track are combined to compute particle identification (PID) confidence levels for the pion, kaon, and proton hypotheses. Tracks are identified as $K$ if the confidence level for the kaon hypothesis is the highest among the three hypotheses of $\pi$, $K$, and $p$. The $\phi$ signal is reconstructed by the invariant mass of the charged kaon pairs candidate with $|M_{K^{+}K^{-}}-m_{\phi}|<18$ MeV/$c^{2}$ and the region of $1055<M_{K^{+}K^{-}}<1127$ MeV/$c^{2}$ is taken as the $\phi$ sideband, where $m_{\phi}$ is the $\phi$ known mass~\cite{pdg2022}. The range here is determined based on the mass resolution obtained from the signal MC sample.

Photon candidates are identified using showers in the EMC. The deposited energy of each shower must be greater than 25 MeV in the barrel region ($|\!\cos\theta|<0.80$) or greater than 50 MeV in the end-cap region  ($0.86<|\!\cos\theta|<0.92$). To suppress electronic noise and energy depositions not associated with the event, the EMC cluster timing from the reconstructed event start time is further required to satisfy $0 \leq t \leq 700$~ns.

To further suppress the combinatorial background, a four-constraint (4C) kinematic fit imposing four-momentum conservation under the hypothesis of
$\psi(3686)\to \gamma\LLB K^{+}K^{-}$ is performed. The combination with the minimum $\chi^2_{\rm 4C}$ is retained for
further analysis if the number of photons is more than one. Furthermore, the $\chi_{\rm 4C}^{2}$ of the kinematic fit is required to be less than 60 by optimizing the figure of merit $\frac{\varepsilon}{\frac{a}{2}+\sqrt{N_{bkg}}}$~\cite{punzi} via the $\chi_{c2}$ decay mode, where $\varepsilon$ is the detection efficiency, $a=3$ is the expected significance of this measurement process, and $N_{bkg}$ is the number of background events estimated from
the inclusive $\psi(3686)$ MC sample. Additionally, the background events of $\chi_{cJ}\to\Omega^{-}\bar{\Omega}^{+}$ are vetoed by the two-dimensional (2D) mass window, $|M_{\Lambda K^{-}}-m_{\Omega^{-}}|>12$ MeV/$c^2\cup|M_{\bar{\Lambda} K^{+}}-m_{\bar{\Omega}^{+}}|>12$ MeV/$c^2$, where $m_{\Omega^{-}({\bar{\Omega}^{+}})}$ is the known mass of the $\Omega^{-}(\bar{\Omega}^{+})$ baryon~\cite{pdg2022}.

\begin{figure}[htbp]
		
		\begin{minipage}[t]{0.82\linewidth}
		\includegraphics[width=1\textwidth]{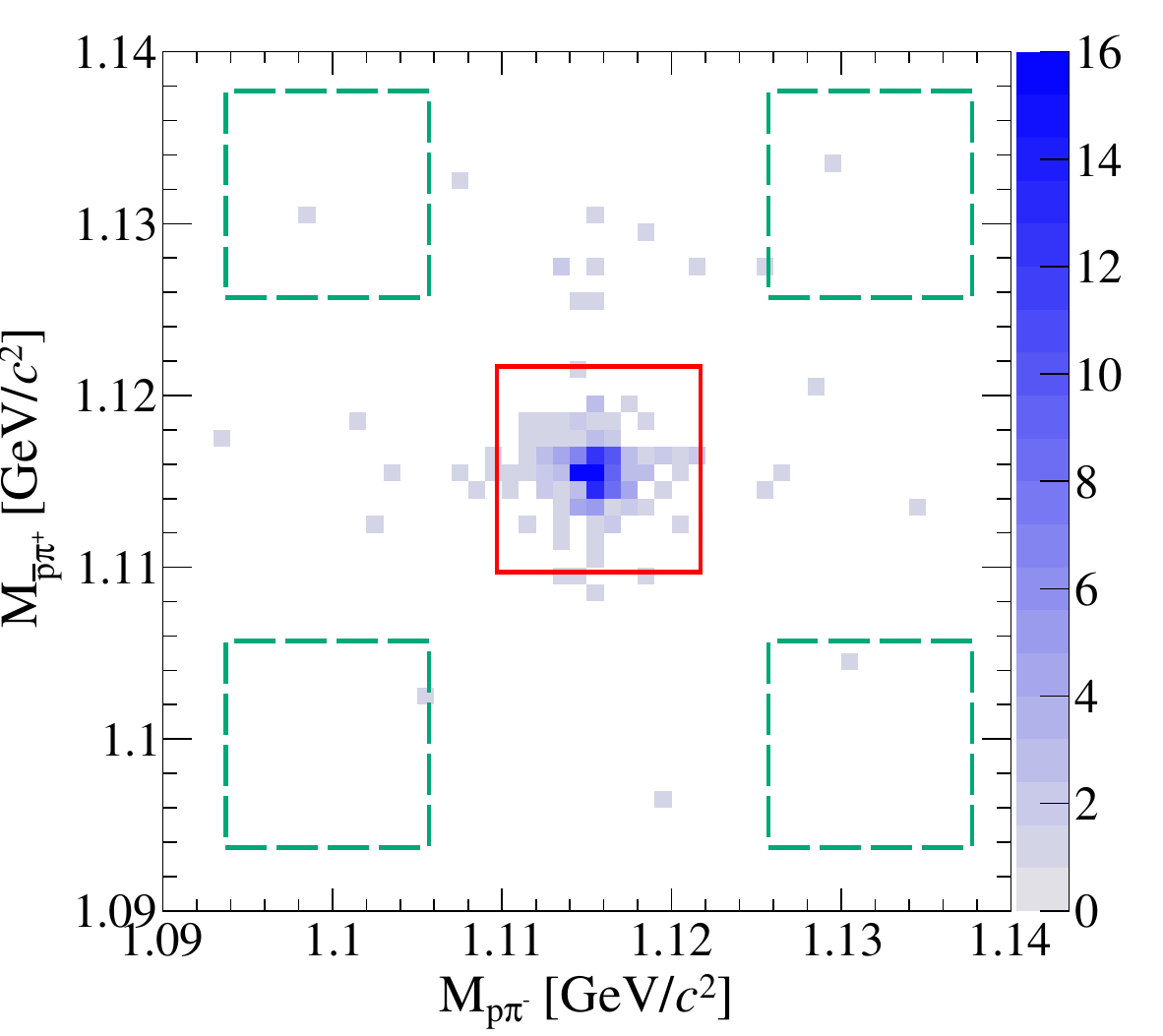}
		\end{minipage}
			
\caption{The 2D distribution of $M_{\bar{p}\pi^{+}}$ versus $M_{p\pi^{-}}$ of the accepted candidates, where the box in red solid lines is the $\LLB$ signal region, and the boxes in green dashed lines are the sideband regions.}
		\label{fig:2D_LLB}
		\end{figure}

Figure~\ref{fig:2D_LLB} shows the 2D distribution of the invariant mass for $p\pi^{-}$ versus $\bar{p}\pi^{+}$ for data after all selection criteria are applied. A clear $\Lambda \bar{\Lambda}$ pair signal can be observed. The one-dimensional (1D) $\Lambda$ and $\bar{\Lambda}$ signal region are defined as $|M_{p\pi^{-}(\bar{p}\pi^{+})}-m_{\Lambda(\bar{\Lambda})}|<6$ MeV$/c^{2}$. The 2D signal region of $\Lambda\bar{\Lambda}$ and its 2D $\Lambda\bar{\Lambda}$ sideband regions are
shown in  Fig.~\ref{fig:2D_LLB}.

Potential remaining background contributions are investigated with the inclusive $\psi(3686)$ MC events, using the event-type analysis tool TopoAna~\cite{zhouxy_topoAna}.  The dominant background comes from the non-resonant $\psi(3686)$ decay of $\psi(3686)\to\gamma\Lambda\bar{\Lambda}\phi$.

To investigate possible continuum background, the same selection criteria are applied to the data samples collected at the center-of-mass energies of 3.650 GeV and 3.682 GeV, corresponding to an integrated luminosity of 454 $\rm{pb}^{-1}$ and 404 $\rm{pb}^{-1}$. No event survives after applying all the selection criteria. Hence, the continuum background is considered to be  negligible.

\section{\label{Sec:BR_determined}Signal yields}
The signal yields of the $\chi_{c0,1,2}\to \Lambda\bar \Lambda\phi$ decays are obtained by performing a simultaneous fit to the $M_{\Lambda\bar{\Lambda}\kk}$ spectra of the events in the $\phi$ signal and sideband regions. In the fit, the signal shapes are described by the MC-simulated shapes convolved with a Gaussian function, and their parameters are floated and shared by all $\chi_{cJ}$ signal contributions.

\begin{figure}[htbp]
\begin{center}
\begin{minipage}[t]{0.9\linewidth}
\includegraphics[width=1\textwidth]{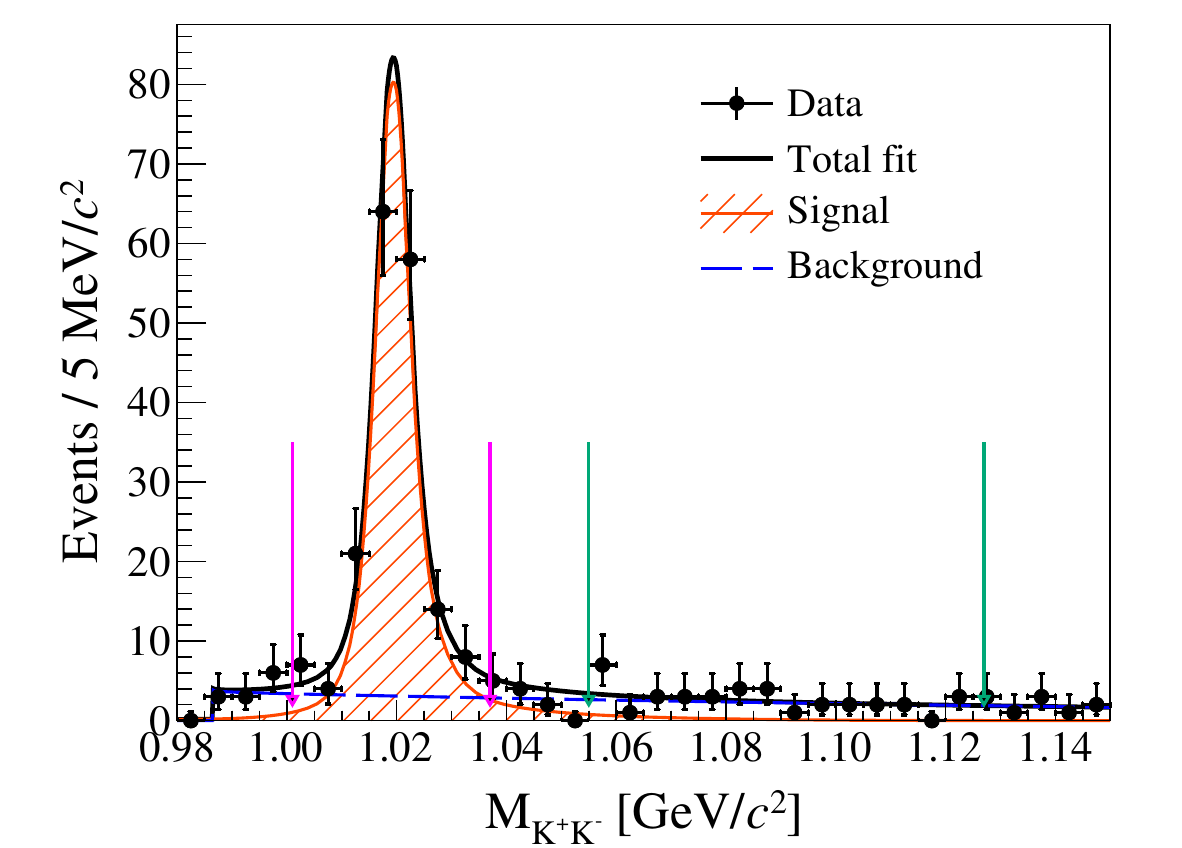}
\end{minipage}%

\caption{Fit to the $M_{K^+K^-}$ distribution of the accepted candidates.
The pink arrows shows the $\phi$ signal region, and the pair of green arrows shows the $\phi$ sideband region. }
\label{fig:mkk}
\end{center}
\end{figure}

\begin{figure*}[htbp]

		\begin{minipage}[t]{0.9\linewidth}
		\includegraphics[width=1\textwidth]{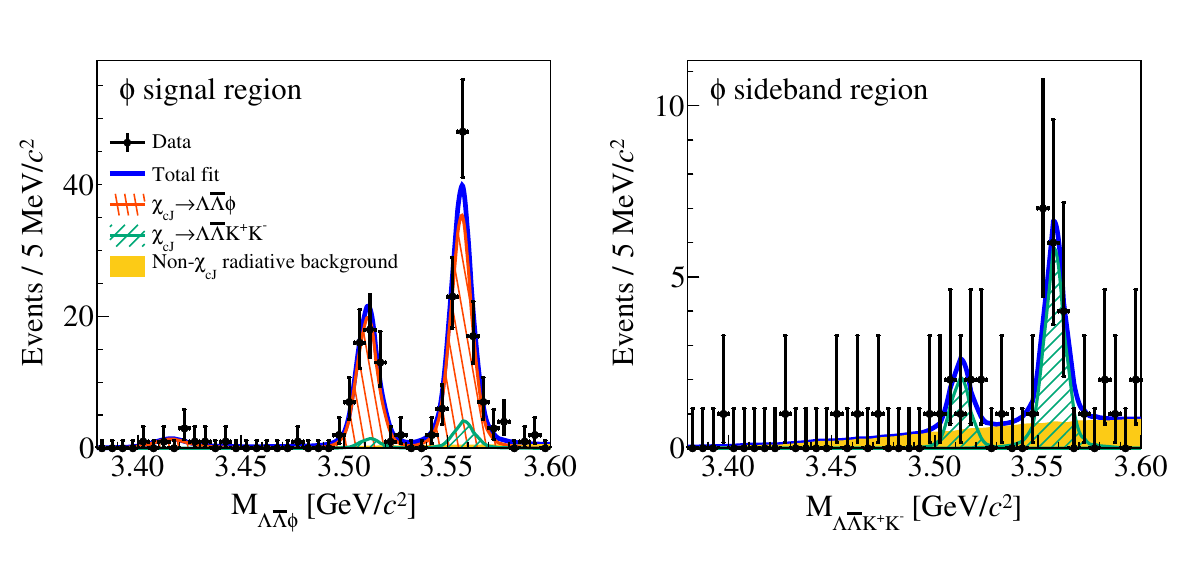}

		\end{minipage}

\caption{Simultaneous fit to the $M_{\Lambda\bar{\Lambda}K^{+}K^{-}}$ distributions in the $\phi$ (left) signal and (right) sideband regions.
}
		\label{fig:simultaneousFit}
		\end{figure*}

We also examine the impact of fake $\Lambda$ and  $\bar{\Lambda}$ candidates.  After considering the number of candidates in the 2D sideband region (denoted as four green boxes in Fig.~\ref{fig:2D_LLB} this background is judged to be negligible.
The non-$\chicJ$ radiative background is described by the MC-simulated shape of the $\psi(3686)\to\gamma\Lambda\bar{\Lambda}\phi$ decay and the number of events is taken as a free parameter.
There is the possibility of $K^+K^-$ background  that does not come from $\phi$ resonance. The shape of this background contribution is derived from the MC sample of $\psi(3686)\to\gamma\chi_{cJ}$, $\chi_{cJ}\to\Lambda\bar{\Lambda}K^{+}K^{-}$, and its yield is treated as a shared parameter for the two modes, with its contribution multiplied by a normalization factor $f_\phi$. The normalization factors, $f_{\phi}$, between the $\phi$ signal and sideband regions are determined to be 0.71 by comparing the numbers of background events in the $\phi$ signal and sideband regions, as shown in Fig.~\ref{fig:mkk}. Specifically, in the fitting of the $K^{+}K^{-}$ invariant-mass spectrum, the signal component is modeled with the MC-simulated shape convolved with a Gaussian function to account for the possible difference in the mass resolution between data and MC simulation. The remaining combinatorial background shape is described by a reverse ARGUS function~\cite{Argus}. The fit result is shown in Fig.~\ref{fig:simultaneousFit} and the obtained signal yields are summarized in Table~\ref{list_summary}.

\begin{table*}[htbp]
\centering\small
\caption{ The signal yields ($N_{\rm obs}^{\chi_{cJ}}$), significances, efficiencies, and the obtained BFs. The first uncertainties are statistical and the second systematic. The significances include systematic uncertainties. }
\label{list_summary}
\begin{tabular}{l |c|c|c}
\hline\hline

&$ {\chi}_{c0} $ & $\chi_{c1} $ & $\chi_{c2} $ \\ \hline

$N_{\rm obs}^{\chi_{cJ}}$ & $7.2\pm3.0$ &  $51.6\pm7.7$ & $94.4\pm10.7$\\
Significance~($\sigma$)  & $4.5$ &  11.3 &  13.0 \\

$\varepsilon(\chi_{cJ}\to\LLB \phi) $~(\%) & 0.45 & 1.61 & 2.54 \\

$\Br(\psi(3686)\to\gamma\chi_{cJ})\cdot\Br\left(\chi_{cJ} \rightarrow \Lambda\bar{\Lambda} \phi\right)$$~(\times10^{-6})$  & $2.92\pm1.22\pm0.19$ &    $5.86\pm0.87\pm0.39$ & $6.79\pm0.77\pm0.35$ \\

$\Br\left(\chi_{cJ} \rightarrow \Lambda\bar{\Lambda} \phi\right)$$~(\times10^{-5})$  & $2.99\pm1.24\pm0.19$ &    $6.01\pm0.90\pm0.40$ & $7.13\pm0.81\pm0.36$ \\

\hline\hline
\end{tabular}
\end{table*}

The significances of $\chi_{c1}\to \Lambda\bar \Lambda\phi$ and $\chi_{c2}\to \Lambda\bar \Lambda\phi$ are determined to be  11.3$\sigma$ and 13.0$\sigma$, respectively, by comparing the difference of likelihoods with and without including each signal in the fit. To estimate the significance of $\chi_{c0}\to \Lambda\bar \Lambda\phi$, we apply another approach due to the low signal yield. We assume that the number of signal and background events in the $\chi_{c0}$ signal region follow a Poisson distribution with mean $n=s+b$~\cite{ZHUYS}, where the signal region is $[m_{\chi_{c0}}-22,~m_{\chi_{c0}}+22]$ MeV/$c^{2}$, with $m_{\chi_{c0}}$  the known $\chi_{c0}$ mass~\cite{pdg2022}, $s$ is the expected number of signal events, while $b$ is the expected number of Poisson distributed background events, and estimated from the
aforementioned fit. Then the $p$-value for the null hypothesis without a resonance ($H_{0}$) is

\begin{align}
\label{significance_Pvalue}
\begin{split}
 p(n_{\rm obs})&=P(n> n_{\rm obs}|H_{0})=\sum\limits_{n = {n_{\rm obs}}}^\infty  {{\textstyle{{{b^n}} \over {n!}}}{e^{ - b}}} \\
& =1 - \sum\limits_{n = 0}^{n_{\rm obs}-1} {{\textstyle{{\frac{b^n}{n!}}}}{e^{ - b}}},\nonumber
\end{split}
\end{align}
where $n_{\rm obs}$ is the number of events observed in the signal region.  All numbers are counted in the signal region where the signal events are expected to appear. The $p$-value is obtained by calculating the probability of the number of background events fluctuating to the number of observed events in the $\chi_{c0}$ signal region. The $p$-value for the $\chi_{c0}\to \Lambda\bar \Lambda \phi$ decay is $7.85\times10^{-6}$, corresponding to a significance of $4.5\sigma$. In determining this significance, the systematic uncertainties are accounted for   by repeating the fits with variations of the signal shape, background shape, and fit range.

\begin{figure*}[htbp]
\begin{center}

\begin{minipage}[t]{0.28\linewidth}
\includegraphics[width=1\textwidth]{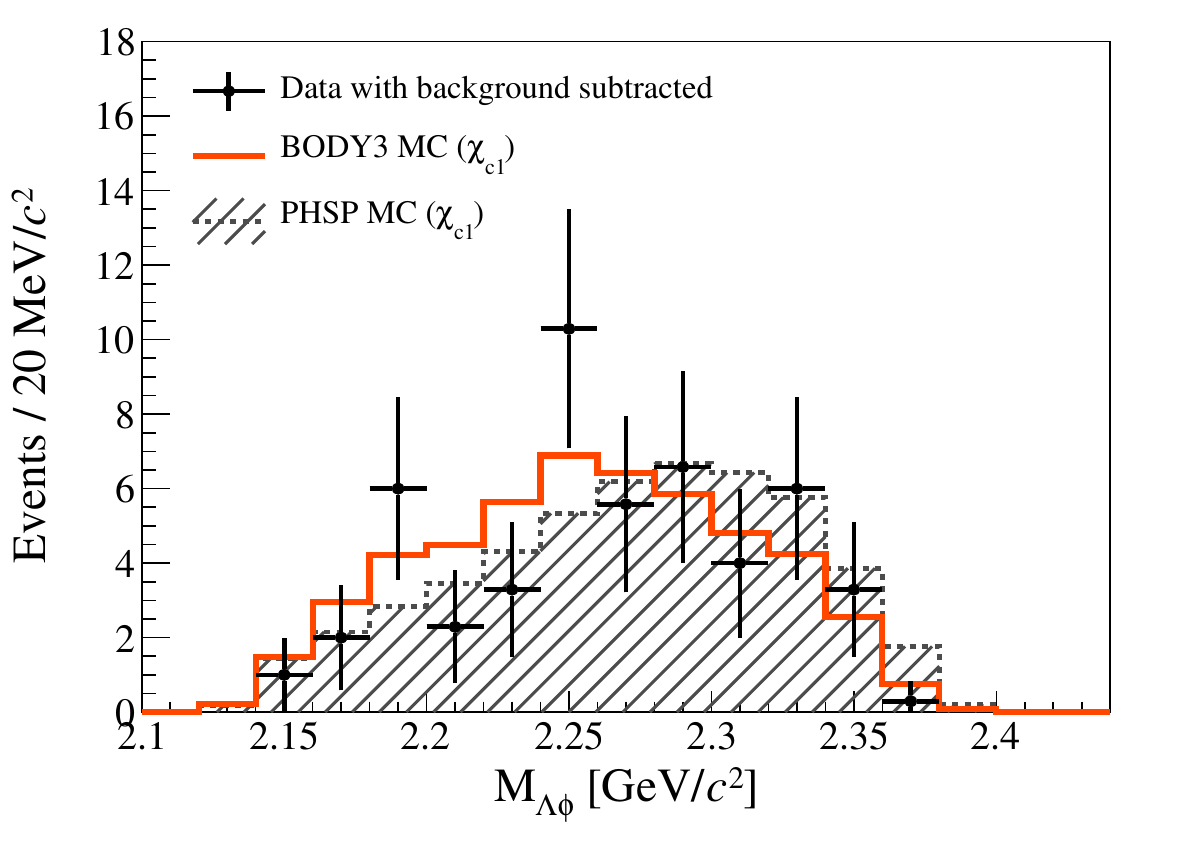}
\end{minipage}
\begin{minipage}[t]{0.28\linewidth}
\includegraphics[width=1\textwidth]{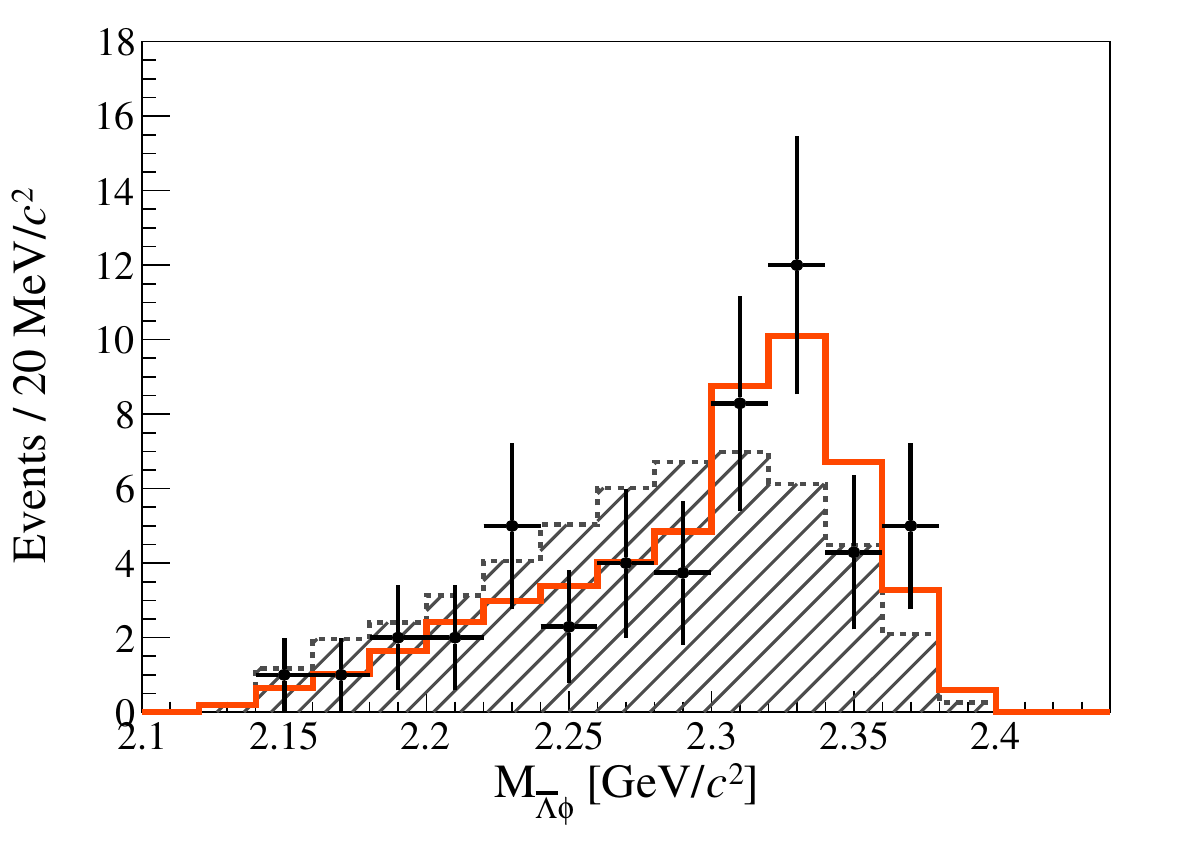}
\end{minipage}%
\begin{minipage}[t]{0.28\linewidth}
\includegraphics[width=1\textwidth]{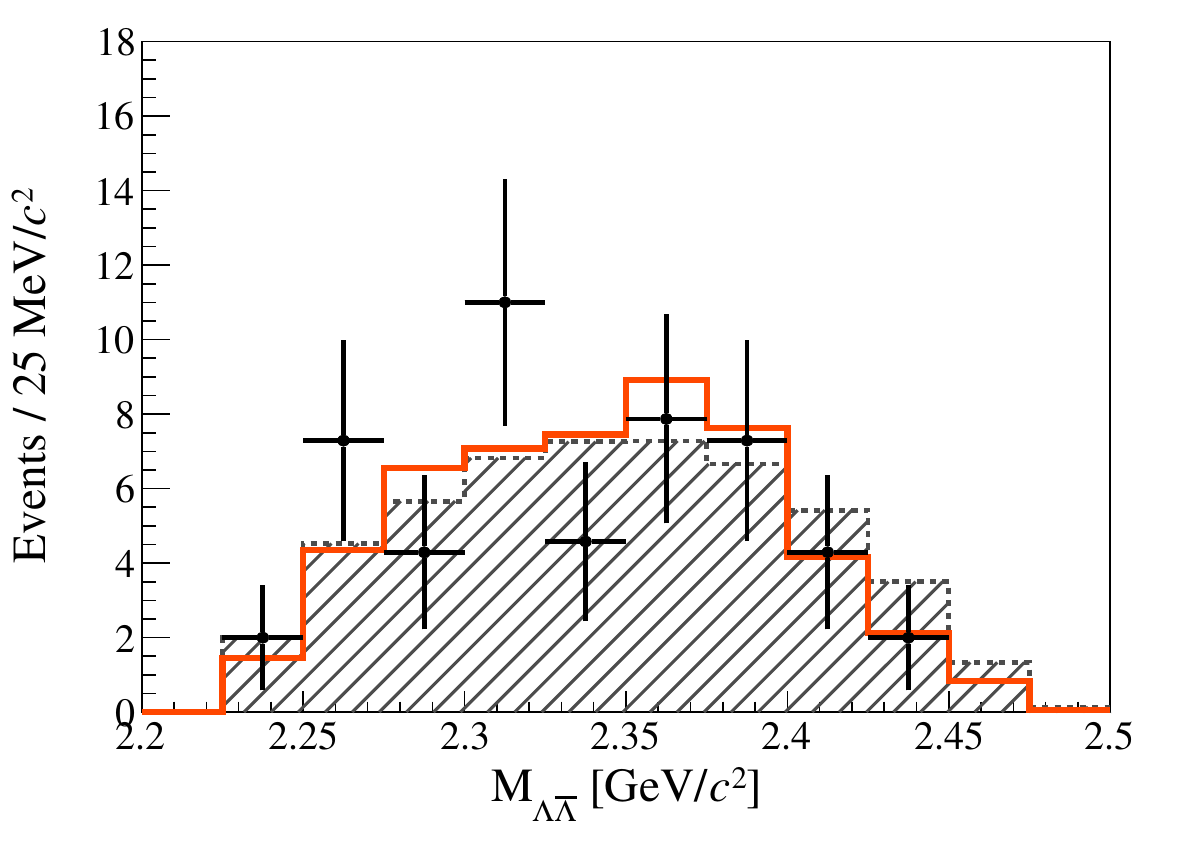}
\end{minipage}

\begin{minipage}[t]{0.28\linewidth}
\includegraphics[width=1\textwidth]{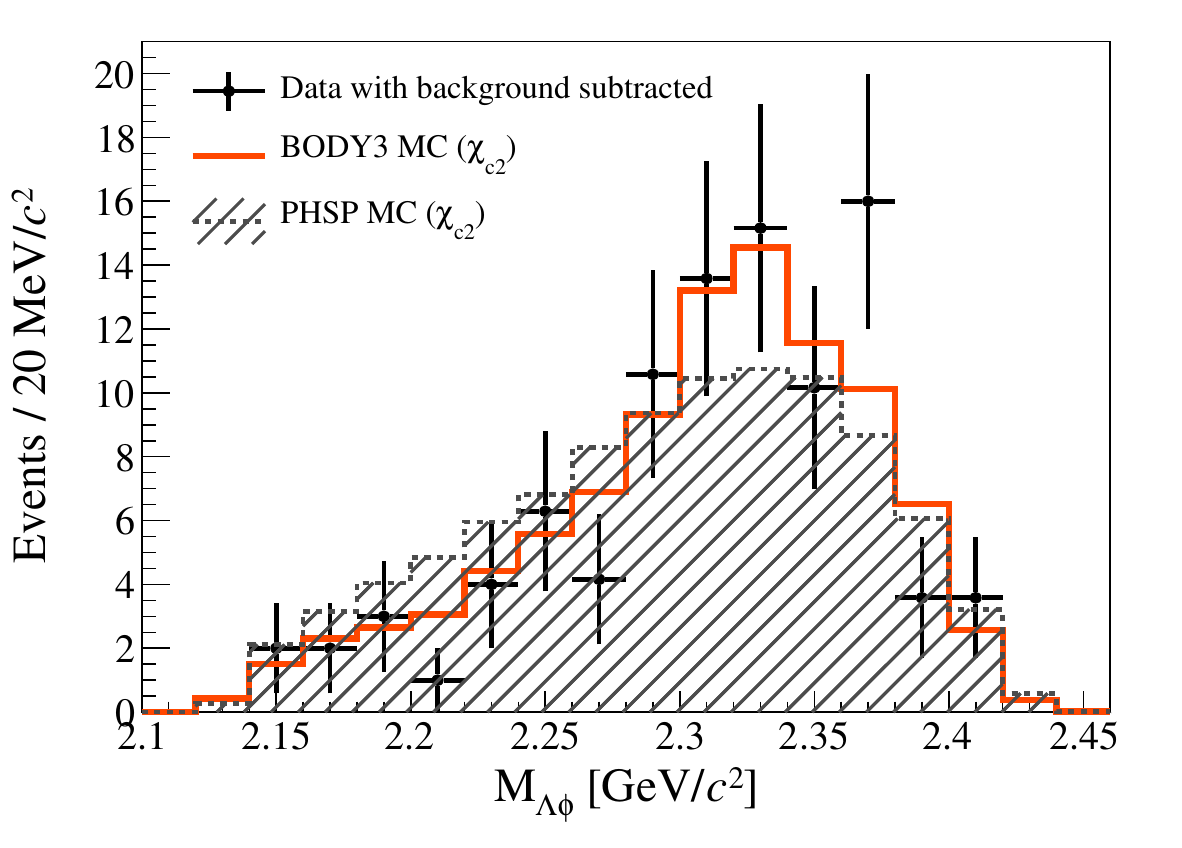}
\end{minipage}
\begin{minipage}[t]{0.28\linewidth}
\includegraphics[width=1\textwidth]{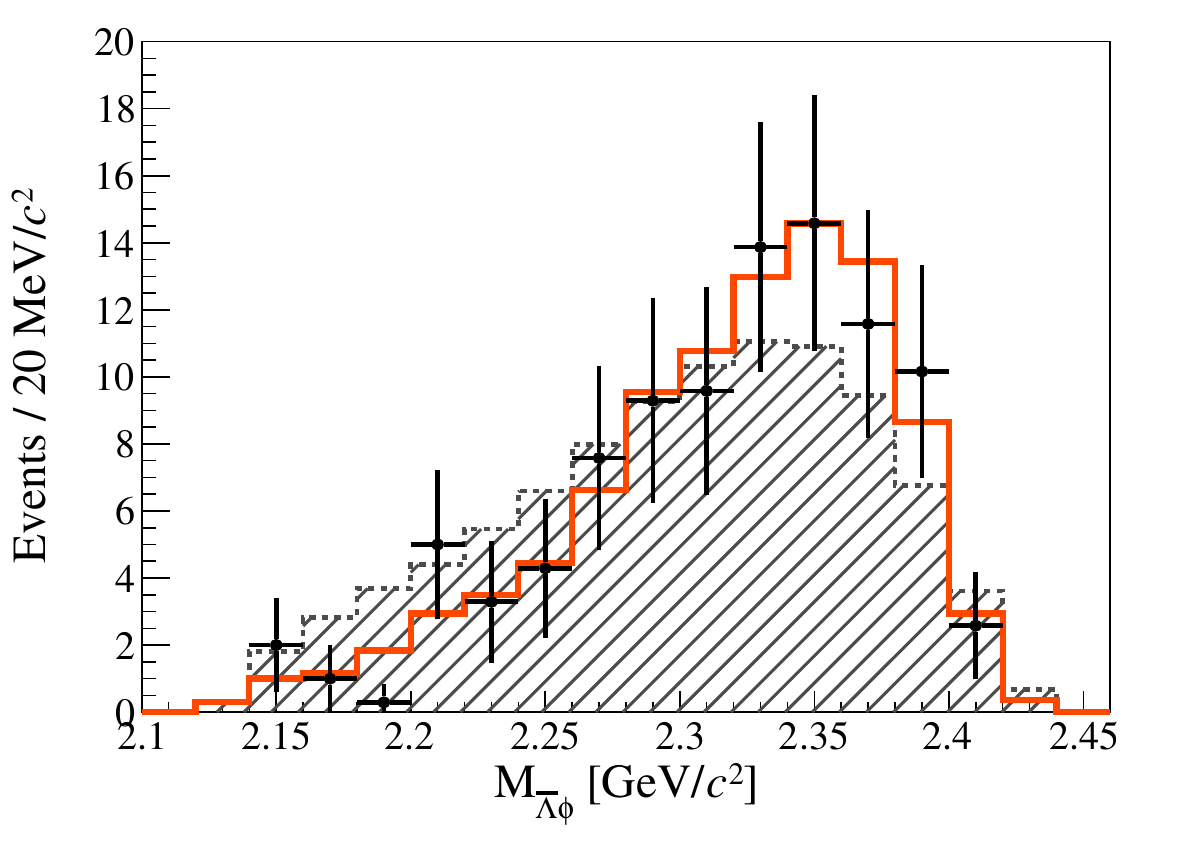}
\end{minipage}%
\begin{minipage}[t]{0.28\linewidth}
\includegraphics[width=1\textwidth]{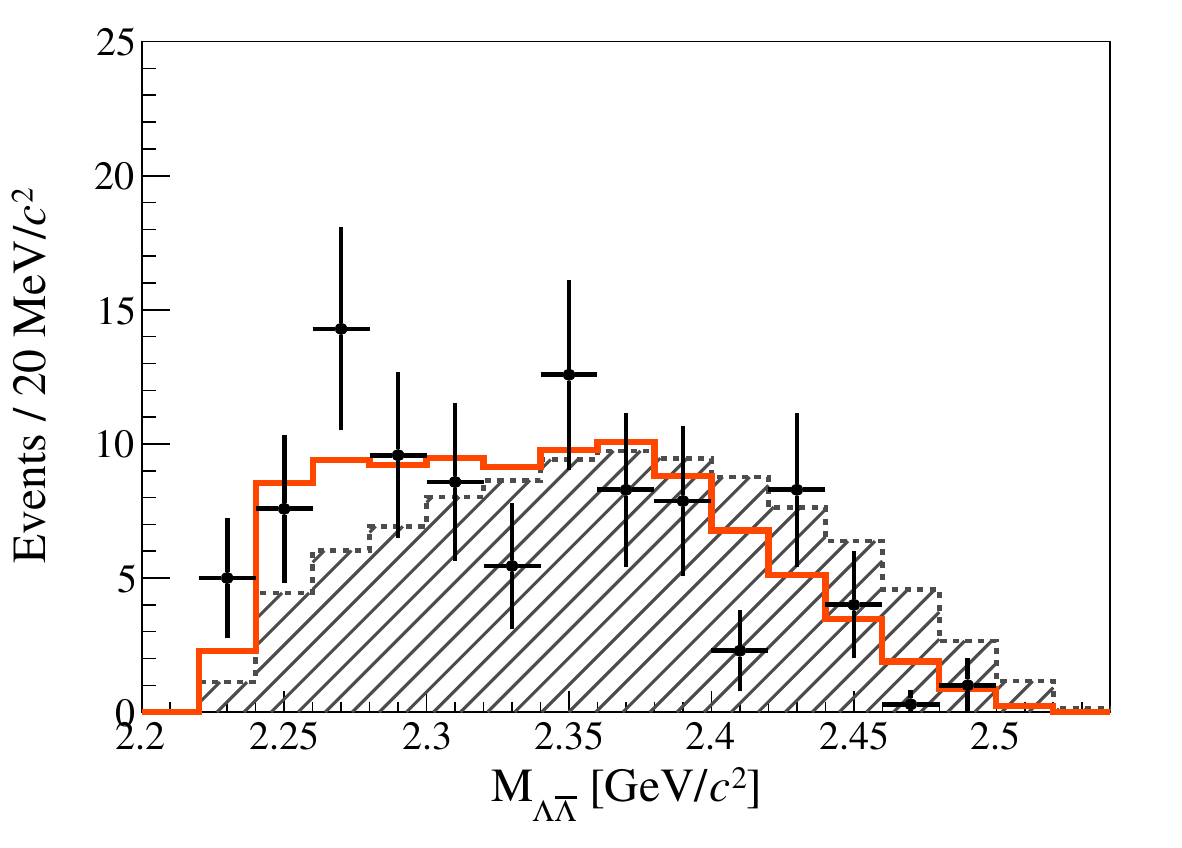}
\end{minipage}

\caption{Invariant-mass distributions of different two-body combinations of the decays of (top row) $\chi_{c1}\to \Lambda\bar\Lambda\phi$ and (bottom row) $\chi_{c2}\to \Lambda\bar\Lambda\phi$.   The data are background subtracted.  Two MC predictions are shown, one based on the PHSP model, the other on the BODY3 model.}
\label{Compare:BODY3_hists_chic12}
\end{center}
\end{figure*}

The existence of potential intermediate states in the $\LLB\phi$ final state are investigated through scrutiny of  the Dalitz plots, apart from in the case of  the $\chi_{c0}$ mode, where the signal yield is too low. Figure~\ref{Compare:BODY3_hists_chic12} shows the invariant masses of different two-body combinations for $\chi_{c1,2}$ signals, after subtracting the background contributions using the normalized sideband events. No obvious structures are observed. Nevertheless, the mass spectra do not agree well with the the signal MC shapes generated with the PHSP model, which will lead to a bias in the efficiency correction made based on this model.  Therefore, the PHSP model is replaced by the modified data-driven generator BODY3~\cite{evtgen_a,evtgen_b}, which was developed to simulate different intermediate states in data for a given three-body final state. The Dalitz plot of $M_{\Lambda\phi}^{2}$ versus $M_{\bar{\Lambda}\phi}^{2}$ found in data, including a binwise correction for backgrounds and efficiencies, is taken as an input to the BODY3 generator. The updated data-MC comparisons based on the BODY3 signal MC samples are shown in Fig.~\ref{Compare:BODY3_hists_chic12}, where the data-MC agreement is improved compared to the PHSP model.

\section{Numerical results}

The BFs of $\chi_{cJ} \to \Lambda\bar{\Lambda}\phi$ are calculated by

\begin{equation}
\Br({\chi _{cJ}} \to \Lambda \bar \Lambda \phi ) = {{{  N_{\rm obs}^{\chi_{cJ}}  \over {N_{\psi \left( {3686} \right)} {\prod _i}  {\Br_i} \cdot \varepsilon(\chi_{cJ}\to\LLB \phi) }}}}   ,
\end{equation}
where $N_{\rm obs}^{\chi_{cJ}}$ is the $\chi_{cJ}$ signal yield, ${\prod _i} {\Br_i}$ is the product BFs of $\psi(3686)\to\gamma\chi_{cJ}$, $\Lambda \rightarrow p \pi$, and $\phi \rightarrow K^{+} K^{-}$, as taken from the PDG~\cite{pdg2022}, $N_{\psi(3686)}$ is the total number of $\psi(3686)$ events~\cite{psip_num_2021}, and $\varepsilon(\chi_{cJ}\to\LLB \phi)$ is the detection efficiency. The measured BFs of each decay mode are summarized in Table~\ref{list_summary}.

\section{Systematic Uncertainties}\label{sec:sysU}
Several sources of systematic uncertainty have been considered and are detailed below.

\begin{itemize}

\item[(i)]{{\bf Kaon tracking:}}
The systematic uncertainty associated with the tracking efficiency is estimated with a control sample of $J/\psi  \to K_S^0{K^ \pm }{\pi ^ \mp }$ decays, and determined to be  $1.0\%$ \cite{Ablikim:2011kv} for each kaon.

\item[(ii)]{\bf Kaon PID:} The systematic uncertainty associated with the kaon-PID efficiency is determined to be 1\%~\cite{Ablikim:2011kv} for each kaon, based on the same sample used to estimate the kaon tracking efficiency.

\item[(iii)]{{\bf Photon  reconstruction:}}
The systematic uncertainty arising from the knowledge of  the photon reconstruction efficiency is assigned to be 1.0\%~\cite{Ablikim:2010zn}, from studies of  the control sample of $J/\psi \to \pi^{+}\pi^{-}\pi^{0}$ decays.

\item[(iv)]{\bf 4C kinematic fit:} To assign the systematic uncertainty related to the 4C kinematic fit,  control samples of $\psi(3686)\to\pi^{+}\pi^{-}J/\psi, J/\psi\to\Lambda\bar{\Lambda}$ and $\psi(3686)\to\eta J/\psi,J/\psi\to\Lambda\bar{\Lambda}\pi^{+}\pi^{-}$, events which have similar topologies as the signal modes are employed. The efficiency of the 4C kinematic fit is defined as the ratio of the signal yields with and without the same $\chi^2$ requirement as the signal channel. The larger difference in efficiencies between data and MC simulation of the two control samples, 1.3\%, is taken as the systematic uncertainty.

\item[(v)]{\bf $\Lambda (\bar{\Lambda})$ reconstruction:}
The combined efficiency including proton/anti-proton and charged pion tracking, and $\Lambda$ and $\bar{\Lambda}$ reconstruction are studied with a control sample of $J/ \psi \rightarrow pK^{-} \bar{\Sigma}^{0}(\rightarrow \gamma \bar{\Lambda} )+ c.c. $ decays. The differences in reconstruction efficiencies between data and MC simulation, $1.8\%$ per $\Lambda$ and $1.5\%$ per $\bar{\Lambda}$, are taken as the corresponding systematic uncertainties.

\item[(vi)]{\bf Signal yield determination}
  \begin{itemize}
  \item[\textbullet] {\bf Mass window:}
  The systematic uncertainties associated with each mass window are estimated by varying each mass window by one standard deviation of the corresponding mass resolution. The larger change of the re-measured BF is taken as the corresponding systematic uncertainty.

  \item[\textbullet]  {\bf Fit range:}
To examine potential systematic uncertainty associated with the choice of  fit range, we perform a Barlow test~\cite{barlow_test} to examine the significance of deviation ($\zeta$) between the baseline fit and the systematic test, defined as
\begin{equation}
    \zeta=\frac{\left|V_{\text {nominal }}-V_{\text {test }}\right|}{\sqrt{\left|\sigma_{V \text { nominal }}^2-\sigma_{V \text { test }}^2\right|}},
\end{equation}
where $V$ is the measured BF and $\sigma_V$ is the statistical uncertainty of $V$. The $\zeta$ distribution is obtained by varying the  fit range  ten times, by shrinking or enlarging the interval (3390, 3590) MeV/$c^2$ to (3370, 3610) MeV/$c^2$, with a step of 2~MeV/$c^2$. After these tests, the associated systematic uncertainty  is found to be negligible since the $\zeta$ distribution shows no significant deviation.

\item[\textbullet]  {\bf Signal shape:}
 To assess the systematic uncertainty due to the choice of signal shape, we use an alternative Breit-Wigner (BW) function $BW(M_{\Lambda\bar{\Lambda}\phi})\times E_{\gamma}^{3}\times D(E_{\gamma})$ to describe the signal distribution, where $BW(M_{\Lambda\bar{\Lambda}\phi})=((M_{\Lambda\bar{\Lambda}\phi}-m_{\chi_{cJ}})^2+\frac{1}{4}\Gamma_{\chi_{cJ}}^2)^{-1}$~\cite{three_body_B} is the nonrelativistic BW function with width $\Gamma_{\chi_{cJ}}$ and  mass $m_{\chi_{cJ}}$  fixed to their individual PDG values~\cite{pdg2022}; $E_{\gamma} = (m_{\psi(3686)}^{2}-M_{\Lambda\bar{\Lambda}\phi}^{2})/2m_{\psi(3686)}$ is the energy of the transition photon in the $\psi(3686)$ rest frame; and $D(E_{\gamma})$ is a damping factor that suppresses the divergent tail due to $E_{\gamma}^{3}$.  This damping factor is described by $D(E_{\gamma})=$exp$(-E_{\gamma}^{2}/8\beta^{2})$ with $\beta$ constrained to the CLEO measurement $(65.0 \pm 2.5)$ MeV~\cite{signal_shape}. The difference in the  signal yields between fits with the two different signal functions is taken as the systematic uncertainty.

  \item[\textbullet]  {\bf Background shape:}
The systematic uncertainties due to the background shape are estimated by replacing the MC-simulated shape of $\psi(3686) \to \gamma \Lambda\bar{\Lambda}\phi$ with a second-order polynomial function. The change of the fitted signal yield is taken as the systematic uncertainty.

\item[\textbullet]  {\bf Normalization factor:}
The systematic uncertainty of the normalization factor of the $\phi$ sideband, $f_{\phi}$, is estimated by varying the sideband region by $\pm 1\sigma$, where $\sigma$ denotes the mass resolution. The largest differences of the BFs from the baseline results are assigned as the corresponding systematic uncertainties.
     \end{itemize}

\item[(vii)]{\bf MC model:}
The systematic uncertainty due to the MC model is estimated by varying the bin size by $\pm25\%$, and varying the number of background events by one standard deviation in the input Dalitz plot in the BODY3 generator, under the assumption that the background satisfies a Poisson distribution. Combining the results from the two sources, the larger difference relative to the baseline efficiency is used to determine the systematic uncertainty.
\item[(viii)]{\bf Input BFs:}
The uncertainties of the BFs of $\psi(3686)\to\gamma\chi_{c0}$, $\psi(3686)\to\gamma\chi_{c1}$, $\psi(3686)\to\gamma\chi_{c2}$, $\Lambda\to p\pi$, and $\phi\to K^{+}K^{-}$ taken from the PDG~\cite{pdg2022} are 2.0\%, 2.5\%, 2.1\%, 1.6\% and 1.0\%, respectively.

\item[(ix)]{\boldmath\bf $N_{\psi(3686)}$}:
The uncertainty on the value of the total number of $\psi(3686)$ events, determined with inclusive hadronic $\psi(3686)$ decays, is 0.5\%~\cite{psip_num_2021}.
\end{itemize}

All the systematic uncertainties are assumed to be independent of each other and combined in quadrature to obtain the overall systematic uncertainty as listed in Table~\ref{list_sys}.

\begin {table}[htbp]
\begin{center}
\renewcommand\arraystretch{1.2}
{\caption{Systematic uncertainties (\%) in the BF measurements, where `neg.' indicates that  the associated systematic uncertainty is  negligible, and the dash indicates that the systematic uncertainty is not applicable.}
\label{list_sys}}
\begin {tabular}{l c c  c}\hline\hline

Source & $\chi_{c0}$ & $\chi_{c1}$  & $\chi_{c2}$  \\   \hline
	
Kaon tracking    &     2.0      &  2.0   & 2.0                      \\
 Kaon PID         &     2.0     &  2.0   & 2.0 	                 \\
	Photon reconstruction   &     $1.0$     &  $1.0$        &  1.0            \\
	4C kinematic fit  &     $1.3$     &  $1.3$    & 1.3                        \\
	$\Lambda$ reconstruction & 1.8  & 1.8& 1.8   \\
	$\bar{\Lambda}$ reconstruction & 1.5 & 1.5& 1.5                  \\
	Mass window       &  1.0    &  0.4  & 1.0                     \\
    Fit range	&  neg.     &  neg.      &        neg.                  \\
    Signal shape	& 3.1      &  $3.1$ & 0.5             \\
    Background shape&  2.8 & $0.8$ &0.3                                  \\
Normalization factor       &    0.3     &  0.2  & 0.3                      \\

    MC model &   $-$   &  $2.6$    &$0.5$                 \\
   Input BFs &  2.8  &  3.1 & 2.8   \\
$N_{\psi(3686)}$  & 0.5 &  0.5 & 0.5 \\  \hline
	 Total & 6.5 &  6.6 & 5.1         \\

\hline
\hline
\end{tabular}

\end{center}
\end{table}

\section{Summary}

By analyzing $(2712.4 \pm 14.3) \times 10^{6}$ $\psi(3686)$ events, we observe or find evidence of the decays of $\chi_{c0,1,2}\to \Lambda\bar \Lambda \phi$ for the first time, with significances of $4.5\sigma$, $11.3\sigma$, and $13.0\sigma$, respectively. We determine their decay BFs to be $\Br(\chi_{c0}\to\Lambda\bar{\Lambda}\phi)=(2.99\pm1.24\pm0.19) \times 10^{-5}$, $\Br(\chi_{c1}\to\Lambda\bar{\Lambda}\phi)=(6.01\pm0.90\pm0.40)\times 10^{-5}$ and $\Br(\chi_{c2}\to\Lambda\bar{\Lambda}\phi)=(7.13\pm0.81\pm0.36) \times 10^{-5}$, where the first uncertainties are statistical and the second systematic. No obvious enhancement near the
$\Lambda\bar{\Lambda}$ production threshold is found. No obvious excited $\Lambda$ state is found in the $M_{\Lambda \phi}$ or $M_{\bar{\Lambda}\phi}$ spectra, either.

\section{ACKNOWLEDGMENTS}
 The BESIII Collaboration thanks the staff of BEPCII and the IHEP computing center for their strong support. This work is supported in part by National Key R\&D Program of China under Contracts Nos. 2020YFA0406300, 2020YFA0406400; National Natural Science Foundation of China (NSFC) under Contracts Nos. 11635010, 11735014, 11835012, 11935015, 11935016, 11935018, 11961141012, 12022510, 12025502, 12035009, 12035013, 12061131003, 12150004, 12192260, 12192261, 12192262, 12192263, 12192264, 12192265, 12221005, 12225509, 12235017, 12275058, 12275067; Program of Science and Technology Development Plan of Jilin Province of China under Contract No. 20210508047RQ and 20230101021JC; Natural Science Foundation of Henan Province (Grant No. 225200810030), and Excellent Youth Foundation of Henan Province (Grant No. 212300410010); the Chinese Academy of Sciences (CAS) Large-Scale Scientific Facility Program; the CAS Center for Excellence in Particle Physics (CCEPP); Joint Large-Scale Scientific Facility Funds of the NSFC and CAS under Contract No. U1832207; CAS Key Research Program of Frontier Sciences under Contracts Nos. QYZDJ-SSW-SLH003, QYZDJ-SSW-SLH040; 100 Talents Program of CAS; The Institute of Nuclear and Particle Physics (INPAC) and Shanghai Key Laboratory for Particle Physics and Cosmology; European Union's Horizon 2020 research and innovation programme under Marie Sklodowska-Curie grant agreement under Contract No. 894790; German Research Foundation DFG under Contracts Nos. 455635585, Collaborative Research Center CRC 1044, FOR5327, GRK 2149; Istituto Nazionale di Fisica Nucleare, Italy; Ministry of Development of Turkey under Contract No. DPT2006K-120470; National Research Foundation of Korea under Contract No. NRF-2022R1A2C1092335; National Science and Technology fund of Mongolia; National Science Research and Innovation Fund (NSRF) via the Program Management Unit for Human Resources \& Institutional Development, Research and Innovation of Thailand under Contract No. B16F640076; Polish National Science Centre under Contract No. 2019/35/O/ST2/02907; The Swedish Research Council; U. S. Department of Energy under Contract No. DE-FG02-05ER41374.


\end{document}